\documentclass[12pt]{article}
\font\lamsbm=msbm10 scaled \magstep2    
\def\be{\begin{eqnarray}}
\def\ee{\end{eqnarray}}
\begin{document}
\centerline{\large VERTEX RING-INDEXED LIE ALGEBRAS} 

\phantom{aaa} 

\centerline{ \Large David Fairlie$^{\S}$ and Cosmas Zachos$^{\P}$ }  

$^{\S}$Department of Mathematical Sciences, Durham University,
Durham, DH1 3LE, UK  \\
\phantom{a} \qquad   \qquad\qquad{\sl David.Fairlie@durham.ac.uk}

$^{\P}$High Energy Physics Division,
Argonne National Laboratory, Argonne, IL 60439-4815, USA \\
\phantom{a}\qquad  \qquad \qquad  {\sl zachos@anl.gov}

\begin{abstract} 
Infinite-dimensional Lie algebras are introduced, which are
only partially graded, and are specified by indices lying on cyclotomic rings.
They may be thought of as generalizations of the Onsager algebra,
but unlike it, or its $sl(n)$ generalizations, they are not subalgebras 
of the loop algebras associated with $sl(n)$. 
In a particular interesting case associated with $sl(3)$, their indices 
lie on the Eisenstein integer triangular lattice, and these algebras 
are expected to underlie vertex operator combinations in CFT, brane physics, 
and graphite monolayers. 
\end{abstract}
\bigskip

\hrule

\section{The New Algebras}

We briefly introduce a class of infinite-dimensional  vertex-operator 
Lie algebras. They have two indices, one of which lacks conventional grading. 
Instead, its composition motivates placing it on a cyclotomic ring, which 
thus makes it effectively equivalent to a multiplet of integers. We expect 
these algebras to feature in CFT and other areas  of physics with enhanced
symmetry. 

Consider the Lie algebras 
\be
[J^a_m , J^b_n ]  =  J^{a+b} _{m+ \omega ^a n }  - J^{a+b}_{n + \omega ^b m}~,
 \label{new}
\ee
where the indices $a,b,... ,~ m,n,...$ and the parameter $\omega $ 
may be arbitrary, in general.

However, as will become evident, the choice of $\omega $ as an $N$-th  root of 
unity, $\omega^N=1$, hence  $1+\omega+ \omega^2+...+\omega^{N-1}=0$,
 and  $a,b,...$ integers, $m,n,...$ proportional to integers, 
yields by far the most interesting family. In that case, 
the upper indices are only distinct mod$N$,
and the lower indices take values in the cyclotomic integer ring
{\lamsbm Z}$[\omega ]$, namely, 
$r+s \omega + k \omega^2+...+ j\omega^{N-2}$. Note the grading of the 
upper indices, but lack of conventional grading for the lower indices,
in contrast to conventional maximally graded two-index infinite Lie algebras 
such as \cite{ffz,hoppe,saveliev}.
(The Lie algebras introduced here appear distinct from 
those based on affine quasicrystals, associated with $N$-th roots of unity, 
Coxeter groups, and Penrose pentilings \cite{reidun} --- 
but some intriguing connection to these algebras should not be excluded.)

This algebra satisfies the Jacobi identity, and possesses the 
central element $J^0_0=J^{-a}_{-\omega^{-a} m} J^a_m$. 
For the cyclotomic family, ``Casimir invariants" may be written as 
\be
J^0_0 = ( J^a_m )^N,  \label{casimir}
\ee
provided $m=0$ if $a=0$. 

In fact, the above Lie algebra might be constructed from the 
group algebra of associative operators 
\be
J^a_m J^b_n =  J^{a+b} _{m+ \omega ^a n }~, \label{group}
\ee
which satisfy 
\be
(J^a_m J^b_n)  J^c_k = J^a_m (J^b_n  J^c_k). 
\ee
It would be customary in such cases \cite{ffz} to also 
consider the anticommutator of these operators, to produce a partner 
graded Lie algebra,
\be
\{ J^a_m , J^b_n \} =  J^{a+b} _{m+ \omega ^a n }  + J^{a+b}_{n + \omega ^b m}~.
\label{susy} 
\ee

A simple operator realization of this algebra is
\be
J^a_m =   e^{m \exp (x)}~ \omega^{a \partial_x} ,  \label{diffreal}
\ee
as may be checked by virtue of the translation action of
$ \omega^{\partial_x} f(x) = f(x+ \ln \omega ) ~ \omega^{\partial_x}$.  
It is easy to see in this realization that the scale of the $a,b$ 
is fixed, but that of the $m,n$ is labile, as they can be 
rescaled with no change to the structure of the algebra.

A variant rewriting of this realization results from the simplifying 
Campbell-Baker-Hausdorff expansion for the particular operators involved,
\be
J^a_m =   \omega^{a\left ( \partial_x +{m\over \omega^a -1 }\exp (x) \right )}. 
\ee
Equivalently, given oscillator operators, $[\alpha ,\alpha ^\dag ]=1$, 
the above realizations may be written in a form evocative of vertex operators,
\be
J^a_m =   e^{m~ \alpha ^\dag }~ \omega^{a ~ \alpha ^\dag \alpha } = 
\omega^{a\left (\alpha^\dag \alpha+{m\over \omega^a -1}\alpha^\dag  \right )}. 
\ee

In the cyclotomic case, $\omega^N=1$,  $a,b$ are equivalent mod$N$, so  
$a,b,.. =0,1,2,...,N-1$.
The $N=2$ case, $\omega=-1$, is trivial, as the corresponding lower index 
ring is that of the conventional integers, and the resulting 
algebra is essentially the Onsager algebra, a subalgebra of the 
$SL(2)$ loop algebra, discussed in the next section.

As an aside, a less ``asymmetric", albeit more cumbersome 
rewriting of eqn (\ref{group}) might be
\be
V^a _m \equiv J^{2a} _{mw^a} ~, 
\ee
so that eqn (\ref{group}) reads 
\be
V^a_m V^b_n =  V^{a+b} _{\omega^{-b}  m+ \omega^a n } ~.
\ee
Antisymmetrization leads to the corresponding notation for the 
Lie algebra (\ref{new}),
\be
 [ V^a_m,  V^b_n ]=  V^{a+b} _{\omega^{-b}  m+ \omega^a n } -
V^{a+b} _{\omega^{b}  m+ \omega^{-a} n } ~.
\ee

\section{$N=2$ Degenerate Case and the Onsager Algebra}
Onsager, in his celebrated solution of the two-dimensional Ising model 
\cite{onsager}, introduced the integer-indexed infinite-dimensional 
Lie algebra,
\be
[A_m ,A_n]= 4 G_{m-n}, \qquad \qquad 
[G_m ,A_n]= 2( A_{m+n} -A_{n-m}), \qquad \qquad 
[G_m ,G_n]= 0 . \label{ons}
\ee
(Also see \cite{davies,uglov,date}.) Evidently, $G_{-m}=- G_m$. 
A potential central element, $G_0$,
is not generated on the r.h.s. of the algebra. Onsager also 
recognized that his algebra is effectively a 
subalgebra of the $SL(2)$ loop algebra ($SU(2)$ centerless Kac-Moody
algebra in modern conventions). The loop Lie algebra 
consists of three integer-indexed ``towers" of elements, with
\be
[ K^{+}_m , K^{-}_n ]= K^{0}_{m+n} ~,  \qquad  
[ K^{0}_m , K^{\pm}_n ]= \pm K^{\pm}_{m+n}~, \qquad  [ K^{\pm}_m , 
K^{\pm}_n ]= [K^{0}_{m} , K^{0}_{n} ] =0 . \label{loop2}
\ee
Given the linear involutive automorphism of this algebra,
\be
K^{\pm}_m \mapsto K^{\mp}_{-m} ~,  \qquad \qquad 
K^{0}_m \mapsto - K^{0}_{-m} ~,  
\ee  
the Onsager algebra is identifiable with the fixed-point 
subalgebra \cite{uglov}, ie, 
the subalgebra invariant under the automorphism, consisting of 
two ``towers",
\be
A_m=2 \sqrt{2} ( K^{+}_m + K^{-}_{-m}), \qquad \qquad 
G_m= 2(   K^{0}_m -K^{0}_{-m}). 
\ee

It can be checked that for $N=2$, thus $a=0,1$, 
the above algebra (\ref{new}) also contains the Onsager algebra as a 
subalgebra,
\be
A_m= 2 J^1_m ~, \qquad G_m=J^0_m - J^0_{-m}                ~. 
\ee
It can then be seen that a graded extension
of the Onsager algebra of the type (\ref{susy}) is trivial, since 
\be
H_{m-n}\equiv \{ A_m, A_n \}= 4(J^0_m+ J^0_{-m} ), 
\ee
check to be central, ie, they commute with all elements, $A_n,G_n$.

Thus, $J^0_m=-J^0_{-m}+$constant; hence, conversely, requiring a 
trivial graded extension of the Onsager algebra essentially 
amounts to (\ref{group}). (Note from eqn (\ref{casimir}) that $A_m A_m$ 
is not an invariant of the Onsager algebra per se, but only 
upon this further condition, $A_m A_m=4J^0_0$.) 

The realization (\ref{diffreal}) reduces here to
\be
A_m =2 e^{m \exp (x) } (-)^{\partial_x} ,  \qquad \qquad  
G_m=e^{m \exp (x) } -e^{-m \exp (x) } .
\ee
In this realization, the potential candidate for a graded extension,
\be
H_m=4( e^{m \exp (x) } +e^{-m \exp (x) } ),
\ee
manifestly commutes with all elements, $A_n,G_n$ .

An alternate realization in terms of Pauli matrices is  
\be
A_m =2 e^{m \sigma_3 } \sigma_1 ~, \qquad \qquad  
G_m= (e^{m} -e^{-m }) \sigma_3 ~,
\ee
similarly illustrating the triviality of 
$H_m\propto$ {\sf 1\kern -0.36em\llap~1}.

\section{$N=3$ and the Eisenstein Integer Lattice}
For $N=3$, the resulting algebra appears to be new, since, for 
$\omega=e^{2\pi i/3}=-1-\omega^2$, the lower indices are of the form 
$m\equiv k+j\omega$ (with integer $k,j$),  
closing under addition, subtraction, and 
multiplication. These comprise the Euclidean ring {\lamsbm Z}$[\omega ]$
of Eisenstein-Jacobi integers \cite{weisstein}, which define a triangular 
2-d lattice with hexagonal rotational symmetry: 
there are three lines at 60$^\circ$ 
to each other going through each such integer and connecting it to its six 
nearest neighbors, forming  honeycomb hexagons. 

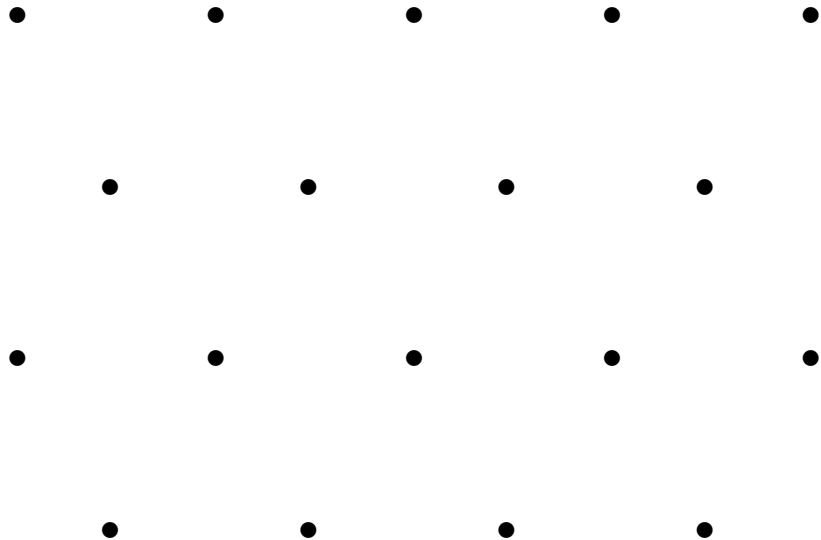
\begin{picture}(380,210)  
\thinlines
\put(110,10){\circle*{6}} 
\put(185,10){\circle*{6}} 
\put(260,10){\circle*{6}} 
\put(335,10){\circle*{6}} 
\put(75,75){\circle*{6}} 
\put(150,75){\circle*{6}} 
\put(225,75){\circle*{6}} 
\put(300,75){\circle*{6}} 
\put(375,75){\circle*{6}} 
\put(110,140){\circle*{6}} 
\put(185,140){\circle*{6}} 
\put(260,140){\circle*{6}} 
\put(335,140){\circle*{6}}
\put(75,205){\circle*{6}} 
\put(150,205){\circle*{6}} 
\put(225,205){\circle*{6}} 
\put(300,205){\circle*{6}} 
\put(375,205){\circle*{6}}  
\end{picture}

This lattice is of utility in 
cohesive energy calculations for monolayer graphite \cite{nanxian}, 
3-state-Potts models associated with WZW CFT models \cite{affleck},
and, perhaps more provocatively, complexifies \cite{conway} to define the 
complex Leech lattice, of significance in string theory, and {\lamsbm Z}$_3$  
orbifolds in CFT \cite{montague}.

Each point on the lattice may be connected to the origin by shifts along the 
$\omega$ root and along the $x$-axis. A 60$^0$ rotation $\omega m$,
on $m\equiv k+j\omega$, for integer coordinates $k,j$, may be represented by 
\begin{equation}
{\mathbf \Omega} \left( \begin{array}{c}
          k\\
          j\\
          \end{array} \right)~ \equiv 
\left( \begin{array}{ccc}
          0&-1\\
          1&-1 \\
            \end{array} \right)\left( \begin{array}{c}
          k\\
          j\\
          \end{array} \right )  , 
\end{equation}
for ${\mathbf \Omega} ^3 =1\kern -0.36em\llap~1 ~$, and 
${\mathbf \Omega} ^2 =-1\kern -0.36em\llap~1 ~ -{\mathbf \Omega}$. Thus, the 
lower indices of the algebra may be considered as a doublet of integers 
composing through this rule.

We care to illustrate this case explicitly to stress the differences from 
conventional loop algebras and $sl(3)$ generalizations of the Onsager algebra.
Instead of the differential realization (\ref{diffreal}), consider 
a faithful representation in terms of $3\times 3$ matrices.
Sylvester's ``nonion"  basis for $GL(3)$ groups \cite{sylvester}, is 
built out of his standard clock and shift unitary unimodular matrices,
\begin{equation}
Q\equiv
\left( \begin{array}{ccc}
          1&0&0  \\
          0&\omega&0  \\
          0&0&\omega^2 \\
            \end{array} \right)~,
\qquad \qquad P\equiv
\left( \begin{array}{ccc}
          0&1&0 \\
          0&0&1 \\
          1&0&0 \\
\end{array} \right)~, 
\end{equation}
so that $Q^3=P^3=1\kern -0.36em\llap~1 ~$.
These obey the braiding identity $PQ =\omega ~ QP$ \cite{sylvester,weyl}. 
For integer indices adding mod 3, the complete set of 
nine unitary unimodular $3\times 3$ matrices
\be
M_{(m_1,m_2)}\equiv \omega ^{m_1 m_2 /2}\, Q^{m_1} P^{m_2},
\ee
where $M^{\dag}_{(m_1,m_2)}=M_{(-m_1,-m_2)}$, and Tr$M_{(m_1,m_2)}=0$,
except for $m_1=m_2=0$ mod3,
suffice to span the group algebra of $GL(3)$. Since
\be
M_{{\bf  m}}M_{{\bf  n}}=\omega^{{\bf  n\times m}/2} M_{{\bf  m+n}}~,
\ee
where ${\bf  m\times n}\equiv m_1 n_2- m_2 n_1$, 
they further satisfy the Lie algebra of $su(3)$ \cite{ffz},
\be
[M_{{\bf  m}},M_{{\bf  n}}]=-2i~\sin\left ( {\pi\over 3}{\bf  m\times n}\right )
~M_{{\bf  m+n}}~.
\ee

It is then simple to realize (\ref{new},\ref{group}) in the 
unimodular $3\times 3$ matrix representation,
\be
J^a_m= e^{mQ} ~ P^a,
\ee
ie, the three ``towers",
\begin{equation}
J^1_m= 
\left( \begin{array}{ccc}
          0&e^m &0  \\
          0&0&e^{m\omega}   \\
          e^{m\omega^2} &0&0 \\
            \end{array} \right),
\qquad J^2_m =
\left( \begin{array}{ccc}
          0&0&e^{m}  \\
          e^{m\omega} &0&0 \\
          0&e^{m\omega^2} &0 \\
\end{array} \right),
\qquad J^0_m =
\left( \begin{array}{ccc}
          e^m &0&0 \\
          0&e^{m\omega} &0 \\
          0&0&e^{m\omega^2}  \\
\end{array} \right) . 
\end{equation}

One may contrast this Lie algebra to not only $su(3)$ loop algebra, but 
also to its subalgebras, such as the the $sl(3)$ generalization of the 
Onsager algebra, introduced by
ref \cite{uglov} and consisting of five towers. Specifically,
the relevant involutive automorphism of $su(3)$ loop algebra, in standard 
Chevalley notation, is
\be
H^{1,2}_m \mapsto - H^{1,2}_{-m}    ~, \qquad 
E^{\pm 1}_m \mapsto E^{\mp 1}_{-m} ~, \qquad 
E^{\pm 2}_m \mapsto E^{\mp 2}_{-m} ~, \qquad 
E^{\pm 3}_m \mapsto -E^{\mp 3}_{-m}~.
\ee
The subalgebra left invariant under this automorphism 
consists of the five towers \cite{uglov},
\be
H^{1,2}_m - H^{1,2}_{-m}~ , \qquad 
E^{1}_m + E^{-1}_{-m} ~, \qquad 
E^{2}_m + E^{-2}_{-m} ~, \qquad 
E^{3}_m - E^{-3}_{-m} ~, 
\ee
or, explicitly,
\begin{equation}
h^{1}_m = \frac{1}{\sqrt{6}}
\left( \begin{array}{ccc}
          e^m-e^{-m} &0 &0  \\
          0&e^{-m}-e^m&0\\
          0 &0&0 \\
            \end{array} \right),
\quad h^{2}_m = \frac{1}{3\sqrt{2}}
 \left( \begin{array}{ccc}
          e^{m}-e^{-m}&0&0\\
          0&e^{m}-e^{-m}&0 \\
          0&0 & 2e^{-m}-2e^{m} \\
\end{array} \right),
\end{equation}
\begin{equation}
e^{1}_m = \frac{1}{\sqrt{3}}
\left( \begin{array}{ccc}
          0&e^m &0  \\
          e^{-m}&0&0  \\
          0 &0&0 \\
            \end{array} \right),
\quad e^{2}_m = \frac{1}{\sqrt{3}}
\left( \begin{array}{ccc}
          0&0&e^{m}  \\
          0&0&0 \\
          e^{-m} & 0&0 \\
\end{array} \right),
\quad e^{3}_m = \frac{1}{\sqrt{3}}
\left( \begin{array}{ccc}
          0&0&0 \\
          0&0& e^m \\
          0&-e^{-m} &0\\
\end{array} \right). 
\end{equation}

\section{General Case: $N>3$, and Quasicrystals}

For higher N, the cyclotomic integer rings {\lamsbm Z}$[\omega ]$  
are less compelling, and are only linked to quasicrystals.
Specifically, the 2-dimensional  complex plane quasilattice 
fills up densely with the set of indices, which fail to close to a 
``sparse'' periodic structure analogous to the 
Eisenstein lattice. A quasicrystal is a higher-dimensional deterministic 
discrete periodic structure whose projection to an embedded ``external space"
(in our case, the complex plane) yields nonperiodic structures of 
enhanced regularity \cite{quasicrystals}.  

For example, for $N=5$, motions are symmetric on a 4-dimensional periodic
lattice, ${\mathbf \Omega} ^5 =1\kern -0.36em\llap~1 ~$, and 
${\mathbf \Omega} ^4 =-1\kern -0.36em\llap~1 ~ -{\mathbf \Omega} 
-{\mathbf \Omega}^2 -{\mathbf \Omega}^3$, with 
\begin{equation}
{\mathbf \Omega} \equiv 
\left( \begin{array}{cccc}
          0& 0& 0& -1\\
          1& 0& 0& -1\\
          0& 1& 0& -1\\
          0& 0& 1& -1\\
            \end{array} \right) , 
\end{equation}
so lower indices may be effectively regarded as a quartet of 
integers---and, likewise, an $N-1$-tuplet of integers for higher $N$. 
However, projected on the actual complex plane,
nearby numbers are not necessarily represented by contiguous points on the 
4-d lattice. 

As indicated at the beginning, there may be links between the present 
algebras over cyclotomic fields  and those on quasicrystals which exhibit a 
five-fold symmetry \cite{reidun}. For $\omega^5  =1$ and  the golden ratio,
$\tau\equiv {1\over 2} (1+\sqrt{5})$, which satisfies $\tau^2=1+\tau$, one
sees that $\tau =-\omega^2 -\omega^3$, since then
$1+\omega+\omega^2+\omega^3+\omega^4 =0$ follows. There is considerable work
\cite{reidun} on algebras defined over such quadratic number fields, 
{\lamsbm Z}$[\tau]=$ {\lamsbm Z}$+${\lamsbm Z}$\tau$, while the associated 
geometric constructions of quasicrystal lattices are available in textbooks
\cite{quasicrystals}. Possibly, detailed investigations of the connection with
algebras defined over the cyclotomic fields will be a fruitful source of
insight. Given the vertex operator realization of the Lie algebras introduced
here and its evocation of coherent states, useful applications in CFT and brane
physics appear likely.

\bigskip 

\noindent{\it This work was supported by the US Department of Energy, 
Division of High Energy Physics, Contract W-31-109-ENG-38.
Discussions with V Armitage and J Hoppe are acknowledged.}

\end{document}